\documentclass[lettersize,journal]{IEEEtran}
\usepackage{amsmath}
\usepackage{amssymb}
\usepackage{graphicx}
\usepackage{cite}
\usepackage{graphicx}
\usepackage{stfloats} 
\usepackage[caption=false,font=footnotesize]{subfig}
\begin{document}
\title{Least-Squares Design of Chromatic Dispersion Compensation FIR Filters Realized with Overlap-Save Processing}%
\author{Oscar Gustafsson,~\IEEEmembership{Senior~Member,~IEEE}, Cheolyong Bae,~\IEEEmembership{Student~Member,~IEEE},\\ and H{\aa}kan Johansson,~\IEEEmembership{Senior~Member,~IEEE}%
\thanks{O. Gustafsson and C. Bae are with the Division of Computer Engineering, Department
	of Electrical Engineering, Link\"oping University, SE-581 83 Link\"oping,
	Sweden. e-mail: oscar.gustafsson@liu.se}
\thanks{H. Johansson is with the Division of Communication Systems, Department
	of Electrical Engineering, Link\"oping University, SE-581 83 Link\"oping,
	Sweden.}
\thanks{This work was supported by the ELLIIT excellence center under project A18: Rational oversampling in coherent optical communication.}
\thanks{Manuscript received \today.}}

\markboth{Submitted to IEEE Photonics Technology Letters, 2023}%
{Gustafsson \MakeLowercase{\textit{et al.}}: Least-Squares Design of CD Compensation FIR Filters Realized in the Frequency Domain}
	\maketitle   
\begin{abstract}
	A design method for chromatic dispersion compensation filters realized using overlap-save processing in the frequency domain is proposed. Based on the idea to use the values that are normally zero-padded, better results than using optimal time-domain design are obtained without any modification of the overlap-save processing complexity. 
\end{abstract}






\section{Introduction}
Chromatic dispersion (CD) in optical fibers causes pulse widening and is one of the more prominent error sources in coherent transmission \cite{Ip2007,Savory2010,Xu2011,Pillai2012}. CD compensation (CDC) also contributes a notable part of the power consumption in a coherent receiver. 

CD is modeled as a non-linear phase allpass frequency response in the fiber as 
\begin{equation}
C\left({{e}}^{j\omega T}\right) = {{{e}}^{-jK(\omega T)^2}},\  K = \frac{D \lambda^2 z}{4 \pi c T^2}, \label{eq:cd}
\end{equation}
where $D$ is the fiber dispersion parameter, $\lambda$ is the wavelength, $z$ is the propagation distance, and $c$ is the speed of light. In this work, we use $\omega T = 2\pi fT$ as ``digital frequency'' with a sampling period of $T$, corresponding to a sampling frequency $f_s = \frac{1}{T}$.

Hence, a CDC filter that approximates the desired frequency response 
\begin{equation}
	H_{\text{des}}(\omega T) = \frac{1}{C\left({{e}}^{j\omega T}\right)} = {{{e}}^{jK(\omega T)^2}}
\end{equation}
should be designed. Only FIR filters are considered here as IIR filters are not suitable because of the inherent limited speed due to their recursive structure. The frequency response of an $L$-tap FIR filter (filter order $L-1$) is 
\begin{equation}
H\left(e^{j \omega T}\right) = \sum_{l=0}^{L-1} h_l e^{-jl \omega T}
\end{equation}
where $h_l$ is the $l$th impulse response coefficient.
 Different design approaches for FIR CDC filters, i.e., determining the values of $h_l$, have been proposed \cite{Savory2008,Eghbali2014,Sheikh2016}. The estimated filter length is given as  \cite{Savory2008}
 \begin{equation}
L = 2\left\lfloor 2K\pi\right\rfloor + 1 = 2\left\lfloor \frac{D \lambda^2 z}{2  c T^2} \right\rfloor + 1. \label{eq:cdclength}
\end{equation}

\begin{figure}%
	\centering
	\includegraphics[scale=0.7]{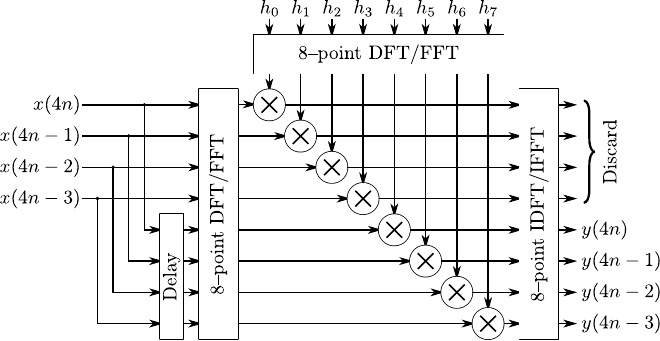}
	\caption{Overlap-save filtering with $M=4$ samples per block using an $N=8$-point DFT/FFT.}
	\label{fig:overlapsaveexample}
\end{figure}

It is common that CDC filters are implemented in the frequency domain\cite{Dinechin2010,Xu2011,Pillai2012,Kovalev2017,Bae2018,Bae2020,Bae2023}, although time-domain implementation has also been proposed \cite{Martins2016,Fougstedt2018,Gustafsson2022}. For correct operation, a scheme such as overlap-add or overlap-save must be used \cite{Harris1987,Blahut2010}, where in each iteration, $M$ samples are processed using an $N$-point discrete Fourier transform (DFT), typically realized using a fast Fourier transform (FFT) algorithm. Overlap-save filtering is illustrated in Fig.~\ref{fig:overlapsaveexample} when processing $M=4$ samples using an $N=8$-point DFT/FFT. 
Normally, the filter length, $L$, is constrained to 
\begin{equation}
	L \leq N -M + 1, \label{eq:Lmax}
\end{equation}
due to the zero-padding required to implement the convolution in the frequency domain.

\begin{figure}%
	\centering
	\subfloat[]{\label{fig:impulseresponse0}
		\includegraphics[scale=0.7]{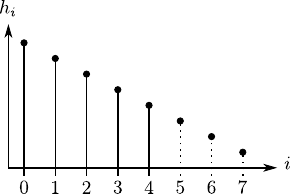}}\hfill
	\subfloat[]{\label{fig:impulseresponse1}\includegraphics[scale=0.7]{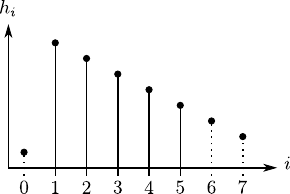}}\hfill
	\subfloat[]{\label{fig:impulseresponse2}\includegraphics[scale=0.7]{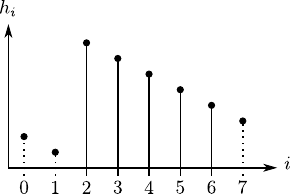}}\hfill
	\subfloat[]{\label{fig:impulseresponse3}\includegraphics[scale=0.7]{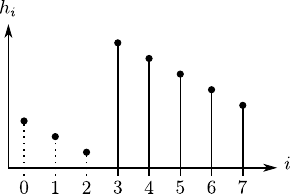}}
	\caption{Associated impulse responses for outputs in Fig.~\ref{fig:overlapsaveexample}: (a) $y(4n)$, (b) $y(4n-1)$, (c) $y(4n-2)$, and (d)  $y(4n-3)$. By zero-extending the five-tap impulse response, the dashed values are zero. In this work, we assign these taps a non-zero value to increase the CDC filtering performance.}
	\label{fig:timefiltering}
\end{figure}

 Consider a impulse response with $h_n = N - n$ applied to the realization in Fig.~\ref{fig:overlapsaveexample}. It can be shown that the four outputs are processed by different, circularly shifted, impulse responses\cite{Johansson2015, Johansson2022}.
These are illustrated in Fig.~\ref{fig:timefiltering}. 
Now, for all outputs to have the same associated impulse response, the dashed impulse response values must be zero%
.
This is the motivation of (\ref{eq:Lmax}), as making the dashed values zero will give the same associated impulse response for all the outputs. However, as will be shown in this work for CDC, and earlier for general filter design \cite{Burel2004,Daher2010}, it is possible to design the impulse response such that the dashed impulse response values are non-zero and benefit from an improved overall performance, despite the filter having a time-varying impulse response and violating (\ref{eq:Lmax}).

In this work, we propose a method for designing CDC filters to be realized using overlap-save filtering in the frequency domain. By utilizing the zero-padding values, we obtain a better CDC filter with the same computational complexity as with zero-padding, supporting longer fibers at the same implementation complexity.

\section{Proposed Design Method}
Introduce a vector
$\mathbf{h} = \begin{bmatrix}
h_0 & h_1 & \dots & h_{N-1}
\end{bmatrix}^T$ corresponding to the effective length-$N$ impulse response, i.e., including the values that are traditionally zero.
Denote a matrix that, when multiplied from the left, circularly shifts a column vector $k$ positions as $\mathbf{S}_k$. Then, the impulse response for output  $y(Mi - m)$, denoted $\mathbf{h}_m$, can be written as $\mathbf{h}_m = \mathbf{S}_m \mathbf{h}$. 

Introduce a length $N$ column vector
\begin{equation}
	\mathbf{d}_{m} = \begin{bmatrix}
		D\left(-\frac{L-1}{2}-m\right) \\
		 D\left(-\frac{L-1}{2}+1-m\right) \\ \vdots \\ D\left(N -1 - m\right)
	\end{bmatrix},
\end{equation}
where 
\begin{IEEEeqnarray}{rCl}
	D(d) & =& \frac{e^{-j\left(\frac{d^2}{4K} + \frac{3\pi}{4} \right)}}{4\sqrt{{\pi }K}} \left({\operatorname{erf}\left({{e^{j\frac{3\pi}{4} } \left(2K\pi -d\right)\over 2\sqrt{K}} }\right)} + \right. \nonumber \\
	& & \left.{{\operatorname{erf}} \left({{e^{j{3\pi \over 4} }\left(2K\pi +d\right)\over 2\sqrt{K}} }\right)}\right),
\end{IEEEeqnarray}
with $K$ from (\ref{eq:cd}) and $\operatorname{erf}$ denoting the error function. Finally, introduce an $N \times N$ symmetric Toeplitz matrix $\mathbf{Q}$ with the element values
\begin{equation}
	Q(n,m) = \begin{cases}
		\frac{\Omega}{\pi}, & m = n\\
		\frac{\sin\left(\Omega(m-n)\pi \right) }{(m - n)\pi}, & m \ne n,\\
	\end{cases}
\end{equation}
where $\Omega$ denotes the bandwidth. Here, only filters with a symmetric bandwidth is considered. An expression for the non-symmetric case is found in \cite{Eghbali2014}.

The optimal impulse response for output $m$ is  \cite{Kidambi1996,Eghbali2014}
\begin{equation}
\hat{\mathbf{h}}_m = \mathbf{Q}^{-1}\mathbf{d}_{m}.
\end{equation}

Considering all $M$ impulse responses simultaneously, the total least-squares error is minimized by solving
\begin{equation}\label{eq:total}
\hat{\mathbf{h}} = \mathbf{R}^{-1}\mathbf{e},
\end{equation}
where
\begin{equation}
\mathbf{R} = \sum_{m=0}^{M-1}\mathbf{S}_m^T\mathbf{Q} \mathbf{S}_m
\end{equation}
and
\begin{equation}
\mathbf{e} = \sum_{m=0}^{M-1}\mathbf{S}_m^T\mathbf{d}_m.
\end{equation}

For the full-band case, i.e., $\Omega = \pi$, it is possible obtain a simpler expression as $\mathbf{Q} = \mathbf{I}$, the identity matrix, and $\mathbf{S}_m^T\mathbf{I} \mathbf{S}_m = \mathbf{I}$ leading to $\mathbf{R} = M\mathbf{I}$ and $\mathbf{R}^{-1} = \frac{1}{M}\mathbf{I}$. Factoring $M$ out from $\mathbf{e}$ and separating it into two vectors give
\begin{equation}
\mathbf{e} = M\begin{bmatrix}
\mathbf{f} \\ \mathbf{g}
\end{bmatrix},
\end{equation}
where $\mathbf{f}$ is of length $L$:
\begin{equation}
\mathbf{f} = \begin{bmatrix}
D\left(-\frac{L-1}{2}\right) \\
D\left(-\frac{L-1}{2} + 1\right) \\
\vdots \\
D\left(\frac{L-1}{2}\right) 
\end{bmatrix},
\end{equation}
and $\mathbf{g}$ is of length $N-L=M-1$:
\begin{equation}
\mathbf{g} = \frac{1}{M}\begin{bmatrix}
\left(M-1\right) D\left(\frac{L-1}{2} + 1\right) + D\left(-\frac{L-1}{2} - M + 1\right) \\
\left(M-2\right) D\left(\frac{L-1}{2} + 2\right) + 2D\left(-\frac{L-1}{2} - M + 2\right) \\
\vdots \\
D\left(\frac{L-1}{2} + M - 1\right) + \left(M-1\right)D\left(-\frac{L-1}{2} - 1\right) 
\end{bmatrix}.
\end{equation}  
Hence, the optimal value for the full-band case is
\begin{equation}
\hat{\mathbf{h}} = \frac{1}{M}\mathbf{I}M\begin{bmatrix}
\mathbf{f} \\ \mathbf{g}
\end{bmatrix} = \begin{bmatrix}
\mathbf{f} \\ \mathbf{g}
\end{bmatrix}. \label{eq:finaldesign}
\end{equation}



It is stressed that although the effective filter length in the proposed approach is $N$ rather than $L=N-M+1$, the DFTs of both cases are of length $N$, so the computational complexities of the realizations are identical.

It should be noted that the terms in both $\mathbf{f}$ and $\mathbf{g}$ are symmetric as $D(d)$ is an even function. Hence, it is possible to rewrite (\ref{eq:finaldesign}) so that fewer variables are used and that the matrix to be inverted is about half the size of the original. Also, it should be noted that although the theoretically optimal solution is given by (\ref{eq:total}), numerical errors may lead to that it is better to use an algorithm to solve the least-square problem using the provided expressions. This is not required for the full-band case, but can improve the numerical accuracy for bandlimited cases.

\section{Results}
\begin{figure*}%
	\centering
	\includegraphics[scale=1.7]{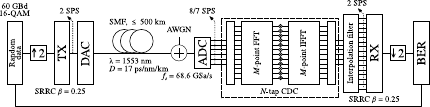}
	\caption{System setup for simulations. Only one polarization shown.}
	\label{fig:system}
	\vspace{-5mm}
\end{figure*}

\begin{figure}%
	\centering

 \includegraphics[]{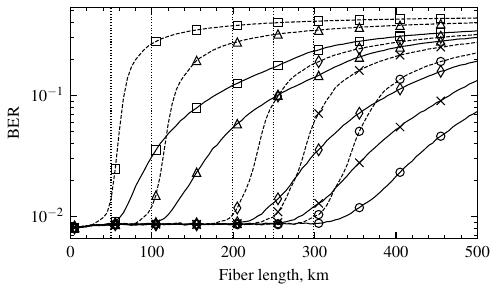}
\vspace{-2mm}
	\caption{BER at different fiber lengths, SNR $= 8$~dB. Dashed: time-domain design \cite{Eghbali2014} with filter length $L$, solid: proposed with FFT length $N = M + L - 1, M=128$. $\Box$:~$L=33$, $\triangle$:~$L=65$ $\diamond$:~$L=129$, $\times$:~$L=161$, and $\circ$:~$L=193$. Vertical dotted lines: fiber lengths from (\ref{eq:cdclength}). \label{fig:length}}
\end{figure}

For the results, a 60~GBd system with 16-QAM modulation and fractional oversampling in the receiver illustrated in Fig.~\ref{fig:system} is considered. This is similar to our earlier works\cite{Fougstedt2020,Bae2020}. As in Bae et al\cite{Bae2020}, no adaptive equalizer is considered. The filter is designed for full bandwidth, i.e., $\Omega = \pi$.
Initially, it is assumed that $M = 128$, i.e., processing 128~samples per clock cycle in a fully parallel implementation of the FFT. Five different values of $N$ are selected, all resulting in efficient FFT implementations\cite{Bae2020} and the fiber length is changed. The bit-error rate (BER) results are shown in Fig.~\ref{fig:length} for an SNR of 8~dB, aiming at an uncoded BER of about $10^{-2}$. It is clear that the proposed filter design technique allows a longer fiber to be used using the same DFT size and processing complexity compared to designing the CDC filter for the time-domain.

\begin{figure}%
	\centering
		\includegraphics[]{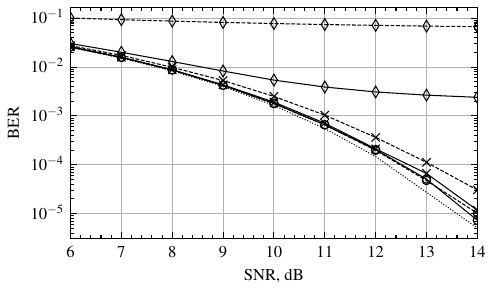}
\vspace{-2mm}
	\caption{BER at 250 km. Dashed: time-domain design\cite{Eghbali2014}  with filter length $L$, solid: proposed with FFT length $N = M + L - 1, M = 128$.  $\diamond$:~$L=129$, $\times$:~$L=161$, and $\circ$:~$L=193$.\label{fig:berat250km2}}
\end{figure}

A traditional BER plot is shown in Fig.~\ref{fig:berat250km2} for three different filter lengths assuming a 250~km long fiber. This corresponds to $L=161$ using (\ref{eq:cdclength}). As seen, for all cases, there is a significant advantage using the proposed design, although for increasing $N$, the benefit decreases.

\begin{figure}%
	\centering
	\includegraphics[]{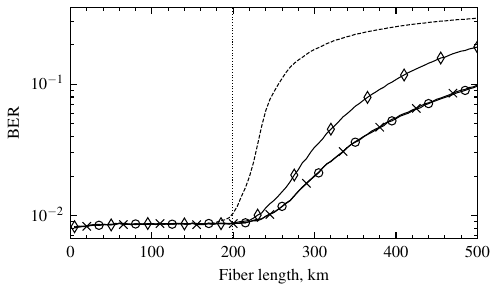}
	\vspace{-2mm}
	\caption{BER at different fiber lengths, SNR $= 8$~dB. Dashed: time-domain design\cite{Eghbali2014} with $L=129$,  $\diamond$: proposed $N=256, M=128$, $\times$: proposed $N=512, M=384$, and $\circ$: proposed $N=1024, M=892$. Vertical dotted line: fiber length from (\ref{eq:cdclength}). \label{fig:length129}}
\end{figure}

To see the impact of only changing the DFT size when realizing filters with the same $L$, the case of $L=129$ is considered, with an estimated maximum fiber length from (\ref{eq:cdclength}) of 200~km. The DFT size $N$ and samples per DFT $M$ are selected as if a filter with $L=129$ is implemented. A common selection in this case is $N = 256$. In addition, $N=512$ and $N=1024$ are considered, which provide readily realizable FFT architectures, but require a non-power-of-two number of input samples per clock cycle\footnote{For $N=512$ and $N=1024$, $96$ and $112$ input samples are processed per clock cycle, respectively, using a 128-parallel FFT implementation \cite{Bae2018}.}. The BER for increasing fiber lengths with an SNR of 8~dB is shown in Fig.~\ref{fig:length129}. It can be seen that longer fibers can be supported at a similar BER level, where $N=512$ and $N=1024$ provide additional improvements over $N=256$.

\begin{figure}%
	\centering
	
	\subfloat[]{\label{fig:BERplot150km}\includegraphics[]{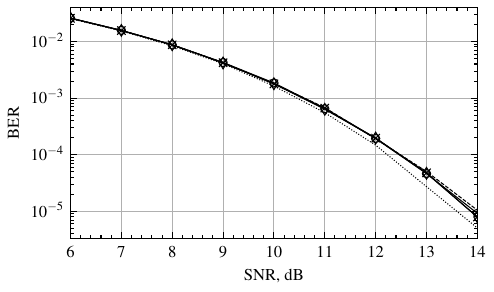}}
	
	
	\subfloat[]{\label{fig:BERplot200km}\includegraphics[]{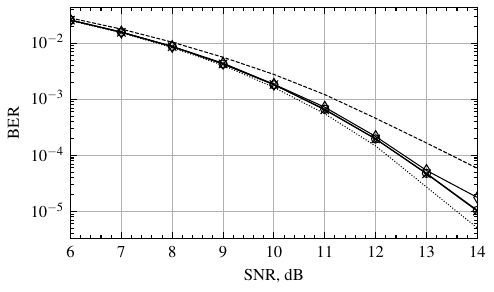}}
	
	
	\subfloat[]{\label{fig:BERplot250km}\includegraphics[]{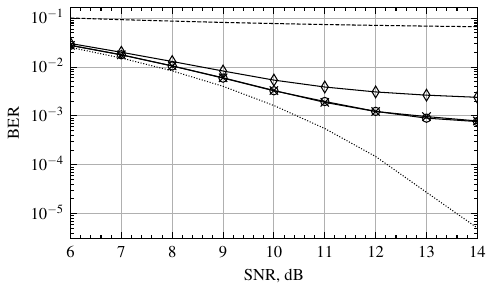}}
	
	\caption{BER at (a) 150~km, (b) 200~km, and (c) 250~km. Dashed: time-domain design with $L=129$\cite{Eghbali2014}, $\diamond$: proposed $N=256,M=128$, $\times$: proposed $N=512, M=384$, $\circ$: proposed $N=1024, M=892$, and dotted: back-to-back.} 
	
\end{figure}

BER results for fibers of 150, 200, and 250~km are shown in Figs.~\ref{fig:BERplot150km} to \ref{fig:BERplot250km}, respectively. It can be seen in Fig.~\ref{fig:BERplot150km} that the proposed approach does not provide any benefits for 150~km. On the other hand, for 200~km, shown in Fig.~\ref{fig:BERplot200km}, there is a significant performance gain using the proposed approach. Finally, for a 250~km fiber, shown in Fig.~\ref{fig:BERplot250km}, it is possible to obtain a much lower BER than for the time-domain design, although for high SNR there is a clear deviation from the theoretical bound. Again, $N=512$ and $N=1024$ provide additional improvements over $N=256$.

\section{Conclusions}

In this work, a chromatic dispersion compensation filter design method for filters realized using overlap-save processing was proposed. Instead of zero-extending the impulse response, non-zero values that optimize the total approximation error in the least-squares sense over all impulse responses simultaneously. The provided simulation results show a significant BER performance improvement obtained without increasing the computational complexity of the overlap-save processing. Hence, the improved CD compensation capabilities come for free if an overlap-save implementation is already used.


\bibliographystyle{IEEEtran}
\bibliography{IEEEabrv,DAabrv,cd}

\begin{thebibliography}{10}
\providecommand{\url}[1]{#1}
\csname url@samestyle\endcsname
\providecommand{\newblock}{\relax}
\providecommand{\bibinfo}[2]{#2}
\providecommand{\BIBentrySTDinterwordspacing}{\spaceskip=0pt\relax}
\providecommand{\BIBentryALTinterwordstretchfactor}{4}
\providecommand{\BIBentryALTinterwordspacing}{\spaceskip=\fontdimen2\font plus
\BIBentryALTinterwordstretchfactor\fontdimen3\font minus
  \fontdimen4\font\relax}
\providecommand{\BIBforeignlanguage}[2]{{%
\expandafter\ifx\csname l@#1\endcsname\relax
\typeout{** WARNING: IEEEtran.bst: No hyphenation pattern has been}%
\typeout{** loaded for the language `#1'. Using the pattern for}%
\typeout{** the default language instead.}%
\else
\language=\csname l@#1\endcsname
\fi
#2}}
\providecommand{\BIBdecl}{\relax}
\BIBdecl

\bibitem{Ip2007}
E.~Ip and J.~M. Kahn, ``Digital equalization of chromatic dispersion and
  polarization mode dispersion,'' \emph{J. Lightw. Technol.}, vol.~25, no.~8,
  pp. 2033--2043, Aug. 2007.

\bibitem{Savory2010}
S.~J. Savory, ``Digital coherent optical receivers: Algorithms and
  subsystems,'' \emph{{IEEE} J. Sel. Topics Quantum Electron.}, vol.~16, no.~5,
  pp. 1164--1179, Sep. 2010.

\bibitem{Xu2011}
T.~Xu, G.~Jacobsen, S.~Popov, M.~Forzati, J.~M{\aa}rtensson, M.~Mussolin,
  J.~Li, K.~Wang, Y.~Zhang, and A.~T. Friberg, ``Frequency-domain chromatic
  dispersion equalization using overlap-add methods in coherent optical
  system,'' \emph{J. Opt. Commun.}, vol.~32, no.~2, pp. 131--135, Jan. 2011.

\bibitem{Pillai2012}
B.~S.~G. Pillai, B.~Sedighi, W.~Shieh, and R.~S. Tucker, ``Chromatic dispersion
  compensation -- an energy consumption perspective,'' in \emph{Proc. Opt.
  Fiber Commun. Conf.}\hskip 1em plus 0.5em minus 0.4em\relax {OSA}, 2012, pp.
  OM3A--8.

\bibitem{Savory2008}
S.~J. Savory, ``Digital filters for coherent optical receivers,'' \emph{Opt.
  Express}, vol.~16, no.~2, pp. 804--817, Jan. 2008.

\bibitem{Eghbali2014}
A.~Eghbali, H.~Johansson, O.~Gustafsson, and S.~J. Savory, ``Optimal
  least-squares {FIR} digital filters for compensation of chromatic dispersion
  in digital coherent optical receivers,'' \emph{J. Lightw. Technol.}, vol.~32,
  no.~8, pp. 1449--1456, Apr. 2014.

\bibitem{Sheikh2016}
A.~Sheikh, C.~Fougstedt, A.~Graell~i Amat, P.~Johannisson, P.~Larsson-Edefors,
  and M.~Karlsson, ``Dispersion compensation {FIR} filter with improved
  robustness to coefficient quantization errors,'' \emph{J. Lightw. Technol.},
  vol.~34, no.~22, pp. 5110--5117, Nov. 2016.

\bibitem{Dinechin2010}
F.~de~Dinechin, H.~Takeugming, and J.-M. Tanguy, ``A {}128-tap complex {FIR}
  filter processing {}20 {Giga-samples/s} in a single {FPGA},'' in \emph{Proc.
  {Asilomar} Conf. Signals Syst. Comput.}\hskip 1em plus 0.5em minus
  0.4em\relax {IEEE}, Nov. 2010, pp. 841--844.

\bibitem{Kovalev2017}
A.~Kovalev, O.~Gustafsson, and M.~Garrido, ``Implementation approaches for
  512-tap 60~{GSa/s} chromatic dispersion {FIR} filters,'' in \emph{Proc.
  {Asilomar} Conf. Signals Syst. Comput.}\hskip 1em plus 0.5em minus
  0.4em\relax {IEEE}, Oct. 2017, pp. 1779--1783.

\bibitem{Bae2018}
C.~Bae, M.~Gokhale, O.~Gustafsson, and M.~Garrido, ``Improved implementation
  approaches for 512-tap 60 {GSa/s} chromatic dispersion {FIR} filters,'' in
  \emph{Proc. {Asilomar} Conf. Signals Syst. Comput.}\hskip 1em plus 0.5em
  minus 0.4em\relax {IEEE}, Oct. 2018, pp. 213--217.

\bibitem{Bae2020}
C.~Bae, P.~Larsson-Edefors, and O.~Gustafsson, ``Benefit of prime factor {FFTs}
  in fully parallel {60 GBaud CDC} filters,'' in \emph{Proc. {OSA} Adv. Photon.
  Congr. -- Signal Process. Photon. Commun.}, Aug. 2020.

\bibitem{Bae2023}
C.~Bae and O.~Gustafsson, ``{FFT}-size implementation tradeoffs for chromatic
  dispersion compensation filters,'' in \emph{Proc. {OSA} Adv. Photon. Congr.
  -- Signal Process. Photon. Commun.}, 2023, p. SpTu3E.1.

\bibitem{Martins2016}
C.~S. Martins, F.~P. Guiomar, S.~B. Amado, R.~M. Ferreira, S.~Ziaie,
  A.~Shahpari, A.~L. Teixeira, and A.~N. Pinto, ``Distributive {FIR}-based
  chromatic dispersion equalization for coherent receivers,'' \emph{J. Lightw.
  Technol.}, vol.~34, no.~21, pp. 5023--5032, Nov. 2016.

\bibitem{Fougstedt2018}
C.~Fougstedt, A.~Sheikh, P.~Johannisson, and P.~Larsson-Edefors, ``Filter
  implementation for power-efficient chromatic dispersion compensation,''
  \emph{{IEEE} Photon. J.}, vol.~10, no.~4, pp. 1--19, Aug. 2018.

\bibitem{Gustafsson2022}
O.~Gustafsson and C.~Bae, ``Shift-and-add realization trade-offs for chromatic
  dispersion compensation {FIR} filters,'' in \emph{Proc. {OSA} Adv. Photon.
  Congr. -- Signal Process. Photon. Commun.}, 2022.

\bibitem{Harris1987}
F.~J. Harris, ``Time domain signal processing with the {DFT},'' in
  \emph{Handbook of Digital Signal Processing}, D.~F. Elliott, Ed.\hskip 1em
  plus 0.5em minus 0.4em\relax San Diego: Elsevier, 1987, pp. 633--699.

\bibitem{Blahut2010}
R.~E. Blahut, \emph{Fast Algorithms for Signal Processing}.\hskip 1em plus
  0.5em minus 0.4em\relax Cambridge University Press, 2010.

\bibitem{Johansson2015}
H.~Johansson and O.~Gustafsson, ``On frequency-domain implementation of digital
  {FIR} filters,'' in \emph{Proc. {IEEE} Int. Conf. Digit. Signal
  Process.}\hskip 1em plus 0.5em minus 0.4em\relax {IEEE}, Jul. 2015, pp.
  315--318.

\bibitem{Johansson2022}
------, ``On frequency-domain impementation of digital {FIR} filters using
  overlap-add and overlap-save techniques,'' 2023, arXiv 2302.08845.

\bibitem{Burel2004}
G.~Burel, ``Optimal design of transform-based block digital filters using a
  quadratic criterion,'' \emph{{IEEE} Trans. Signal Process.}, vol.~52, no.~7,
  pp. 1964--1974, Jul. 2004.

\bibitem{Daher2010}
A.~Daher, E.~H. Baghious, G.~Burel, and E.~Radoi, ``Overlap-save and
  overlap-add filters: Optimal design and comparison,'' \emph{{IEEE} Trans.
  Signal Process.}, vol.~58, no.~6, pp. 3066--3075, Jun. 2010.

\bibitem{Kidambi1996}
S.~S. Kidambi and R.~P. Ramachandran, ``Complex coefficient nonrecursive
  digital filter design using a least-squares method,'' \emph{{IEEE} Trans.
  Signal Process.}, vol.~44, no.~3, pp. 710--713, Mar. 1996.

\bibitem{Fougstedt2020}
C.~Fougstedt, O.~Gustafsson, C.~Bae, E.~B{\"{o}}rjeson, and P.~Larsson-Edefors,
  ``{ASIC} design exploration for {DSP} and {FEC} of {400-Gbit/s} coherent
  data-center interconnect receivers,'' in \emph{Proc. Opt. Fiber Commun.
  Conf.}\hskip 1em plus 0.5em minus 0.4em\relax {OSA}, Mar. 2020, pp. 1--3.

\end{thebibliography}

\end{document}